\documentclass[preprint,3p]{elsarticle}
\usepackage{amssymb}
\usepackage{graphicx}
\journal{Nuclear Physics A}
\begin{document}

\begin{frontmatter}

\title{Phase transitions in neutron star and magnetars and their connection with high energetic bursts in astrophysics}

\date{\today}

\author{Ritam Mallick}
\ead{ritam.mallick5@gmail.com} 
\author{P K Sahu}
\ead{pradip@iopb.res.in}
\address{Institute of Physics, Sachivalaya Marg, Bhubaneswar 751005, INDIA}

\begin{abstract}
The phase transition from normal hadronic matter to quark matter in neutron stars (NS) could give rise to several
interesting phenomena. Compact stars can have such exotic states up to their surface (called strange stars (SS)) or 
they can have quark core surrounded by hadronic matter, known as hybrid stars (HS). As the state of matter of the 
resultant SS/HS is different from the initial hadronic matter, their masses also differ. Therefore, such 
conversion leads to huge energy release, sometimes of the order of $10^{53}$ ergs. In the present work we 
study the qualitative energy released by such conversion. Recent observations
reveal huge surface magnetic field in certain stars, termed magnetars. Such huge magnetic fields
can modify the equations of state (EOS) of the matter describing the star. Therefore, the mass of magnetars
are different from normal NS. The energy released during the conversion
process from neutron magnetar (NM) to strange magnetar/hybrid magnetar (SS/HS) is different from normal NS to SS/HS conversion. 
In this work we calculate the energy release during the
phase transition in magnetars. The energy released during NS to SS/HS conversion exceeds the  
energy released during NM to SM/HM conversion. The energy released during the conversion of NS to 
SS is always of the order of $10^{53}$ ergs. The amount of energy released during such conversion can 
only be compared to the energy observed during the gamma ray bursts (GRB). 
The energy liberated during NM to HM conversion is few times lesser,
and is not likely to power GRB at cosmological distances. However, the magnetars are more likely to lose their
energy from the magnetic poles and can produce giant flares, which are usually associated with magnetars.  
\end{abstract}



\end{frontmatter}

\section{Introduction}

GRB are cosmic gamma ray emission distributed isotropically originating at extra galactic distances.
They are inferred to have energy of the order of $10^{53}$ ergs.
If the beaming and the $\gamma$-ray production of the GRB are quite high, 
then the central engine which powers such conversion can release energy to power GRB at 
cosmological distances. There are various models describing the central engine for GRB. Mergers of two 
NS or a NS and a black hole in a binary \cite{pacz} being some popular models. However, recent calculation of 
Janka \& Ruffert \cite{janka} had shown that the energy released in such neutrino-antineutrino 
annihilation is much smaller to account for the GRB, thought to be 
occurring at large cosmological distances. Therefore, the central engine which powers such huge energies in GRB
still remains unclear.

On the other hand, the conversion of a NS to a QS liberates huge amount of energy, simply 
because the star mass changes during the conversion. Therefore, the mass difference of NS and QS 
manifest itself in the form of energy release. The first calculation of the energy release 
by such conversion was proposed by Alcock et al. \cite{alcock} and Olinto \cite{olinto}.
More detailed similar calculation for the conversion of NS to SS/HS was done by Cheng \& Dai \cite{cheng}, 
Ma \& Xie \cite{ma} and subsequently by many others \cite{bhat1,bhat2,drago,neibergal,herzog,mallick1,pagliara}. 
Almost all of them found the energy released 
to be greater than $10^{51}$ergs, and connected them with the observed gamma ray bursts (GRB).
However, new observation \cite{kulkarni98,kulkarni99} reveals the GRB to occur at cosmological distances. 
Therefore, energies of the order of $10^{51}$ ergs are too low to power GRB at such huge distances. 
A systematic calculation done by Bombaci \& Datta \cite{bombaci}, using various EOS, estimated the
energy released during conversion to be of the order of $10^{53}$ergs. 
Their main assumption was that the initial NS baryonic mass (BM) and the final NS BM
is the same, the BM is conserved during such conversion. Their estimation of the energy
release can power GRB to such large distances and was in agreement with the observed GRB. 
However, during the conversion there can be ejection of matter from the outer layers of the star due 
to several shock bounce. The matter ejected is usually assumed to be low, but Fryer \& Woosley \cite{woosley}
showed that the matter ejected can be as high as $0.2$ solar mass. For such a scenario the assumption of 
baryonic mass conservation does no longer hold. 

The above calculation of energy release depends on the fact that some of the compact stars are
really exotic and quark stars (QS either SS or HS) do really exist.
However, the recent measurement \cite{Demorest10,new2m} found the pulsars to have a 
very high mass, $M\sim 2 $M$_\odot$. Pulsars with such high masses, for the 
first time, has imposed strict constraints in determining the equations of state (EOS) of 
matter describing compact stars. The EOS of quark matter usually has strangeness 
\cite{itoh,bodmer,witten} and therefore, provides an additional 
degree of freedom. This extra degree of freedom softens the EOS, that is, the reduction of
pressure for a given energy density. As a result, it becomes difficult for the quark EOS to 
generate stars with such high mass which satisfies the mass limit set by new measured heavy pulsars. 
However, new calculations have found that the effects due to strong 
interaction, such as one-gluon exchange or color-superconductivity can make the EOS stiffer and increase 
the maximum mass limit of SS and HS \cite{Ruester04,Horvath04,Alford07,Fischer10,Kurkela10a,
Kurkela10b}. Ozel \cite{Ozel10} and Lattimer \cite{Lattimer10} were the first to study the implications 
of the new mass limit for SS and HS within the MIT bag model. Therefore, the conversion
of NS to QS (SS/HS) is still a viable scenario of astrophysical phase transition (PT).

The only other astrophysical events which can come close to the energy budget of GRB are the recently observed
giant flares. Three giant flares, SGR 0526-66, SGR 1900+14 and SGR 1806-20 \cite{flare} have been detected so far. 
The huge amount of energy in these flares can be explained by the presence of a strong surface magnetic field, whose 
strength is estimated to be larger than $10^{14}$G, and such stars are termed magnetars. Simple calculation suggests
that the magnetic fields in the star interior can be few order higher ($\sim 10^{18}$G).
Theoretical calculation also suggests that such strong magnetic field can effects the gross structure of NS. It effects 
the structure of a NS through the metric describing the star \cite{bocquet,cardall} or 
by modifying the EOS of the matter through the Landau quantization of charged particles 
\cite{chakrabarty97,yuan,broderick,chen,wei}. As the EOS gets modified, the mass of magnetar differs from a normal 
pulsar. This would eventually effect the energy released during the conversion of a NM to quark magnetar (QM). 
Therefore, the energy released during the PT of normal pulsars would be quite different from the energy released 
during the PT of magnetars. 

In the present work, we calculate the energy release during the conversion of NS to SS/HS. 
Previous calculation does not include the idea of matter ejection during the conversion process.
In our calculation we assume that the baryonic mass conservation to be true, only after taking into account the mass 
of the matter ejected. Finally, we would study the energy released during the conversion of a NM to
SM/HM. In the next section (Section II), we give the outline of 
the calculations involved in the energy release. In section III, we discuss the EOS. 
We discuss the effect of magnetic field on the EOS in section IV, and the HS is discussed in section V.
The results are discussed in section VI. Finally, in section VII, we summarize our results and draw conclusion from
them.
 
\section{Energy released during conversion}

Earlier calculation of the energy release during the phase transition of a NS to QS
was based on the idea that the baryon number is conserved during the phase transition, therefore, the baryonic mass 
of the star remains same after the conversion. However, there is a change in the proper mass and the 
gravitational mass. The energy released during the conversion is primarily due to the difference in the 
proper and gravitational mass of initial NS and final QS. However, Fryer \& Woosley \cite{woosley}
showed that there may be ejection of matter during the conversion from the outer layer of the star due 
to several shock bounce. Their calculation showed that the matter ejected can be as high as $0.2$ solar mass,
where the baryon number conservation is not maintained. In this work we assume that as the PT
occurs and the transition front travels outwards converting hadronic matter to quark matter. During the PT the star suffers 
several shock bounce and matter is ejected 
primarily from the outer layers of the NS. Matter is ejected before the phase transition front 
reaches the outer layers. Therefore, the matter ejected is normal hadronic matter. 

What happens is that, say the phase transition starts in a NS of baryonic mass $2$ solar mass. The phase transition 
front starts travelling outwards, towards the surface of the NS, converting hadronic matter
to quark matter. During this time several shock bounces happens and matter is ejected from the outer layers of the star.
The ejected matter is still hadronic matter (mostly from the outer nuclear region). We assume the matter ejected is of 
baryonic mass $0.2$ solar mass. Therefore, the baryonic mass of the NS is now $1.8$ solar mass. 
The baryonic mass of the final QS is also $1.8$ solar mass. The energy
released during the conversion is due to the difference in the proper and gravitational mass of the 
intermediate NS and final QS (which have the same baryonic mass of $1.8$ solar mass). The matter ejected does not play 
any role in the energy release because it is ejected from the star before suffers any phase transition. 

The total energy released in the conversion of NS to QS is actually the difference 
between the total binding energy of the QS star (BE(QS)) and the total binding energy of the NS (BE(NS))    
\begin{equation}
 E_T = BE(QS) - BE(NS).
\end{equation}
The total conversion energy can then be written as the sum of gravitational and internal energy change 
\begin{equation}
E_T = E_I  +  E_G.  
\end{equation}
where the gravitational and internal energy are given by
\begin{eqnarray}
E_I =  BE_I(QS) - BE_I(NS) \\ \nonumber
E_G =  BE_G(QS) - BE_G(NS).  
\end{eqnarray}

The total, internal and gravitational energy of conversion can be written in terms of their respective 
gravitational and proper mass (or rest mass)
\begin{eqnarray}
E_T = [M_G(NS) - M_G(QS)]c^2, \\ \nonumber
E_I = [M_P(NS) - M_P(QS)]c^2,  \\ \nonumber
E_G = [M_P(QS) - M_G(QS) - M_P(NS) + M_G(NS)]c^2, 
\end{eqnarray}
where $M_G$ is the gravitational mass, $M_P$ is the proper mass or rest mass of a star.
Keeping the baryonic mass of the star fixed, we calculate the proper and gravitational mass of the star. 
The baryonic mass, proper mass and the gravitational mass of a star can be obtained by solving the structural 
equations for non-rotating compact stars by Tolman-Oppenheimer-Volkoff equations \cite{shapiro}. The baryonic mass,
gravitational mass and proper mass are given by 
\begin{eqnarray}
M_B= \int_0^R  dr 4\pi r^2 \Big[ 1 - {{2Gm(r)}\over{c^2 r}}\Big]^{-1/2}  n(r)m_N, \\ 
M_G = \int_0^R 4 \pi r^2 \epsilon(r) dr, \\ 
M_P =   \int_0^R  dr 4\pi r^2 \Big[ 1 - {{2Gm(r)}\over{c^2 r}}\Big]^{-1/2}  \epsilon(r),
\end{eqnarray}
where $n(r)$ is the number density, $m_N$ is the mass of neutron, $\epsilon(r)$ being the total mass-energy 
density and $m(r)$ the gravitational mass enclosed within a spherical volume of radius $r$. 

\section{Hadronic and quark EOS}

We use the non-linear relativistic mean field (RMF) model with hyperons
to describe the hadronic EOS. 
In this model, the baryons interact with mean meson fields \cite{boguta,glen91,sugahara,sghosh,
schaffner}.

The Lagrangian density having nucleons, baryon octet ($\Lambda,\Sigma^{0,\pm},\Xi^{0,-}$) 
and leptons can be written in the form \cite{ritam2012,ritam2012b}
\begin{eqnarray} 
\label{baryon-lag}   
{\cal L}_H & = & \sum_{b} \bar{\psi}_{b}[\gamma_{\mu}(i\partial^{\mu}  - g_{\omega b}\omega^{\mu} - 
\frac{1}{2} g_{\rho b}\vec \tau . \vec \rho^{\mu})  \nonumber \\ 
& - & \left( m_{b} - g_{\sigma b}\sigma \right)]\psi_{b} + \frac{1}{2}({\partial_\mu \sigma \partial^\mu 
\sigma - m_{\sigma}^2 \sigma^2 } ) \nonumber \\ 
& - & \frac{1}{4} \omega_{\mu \nu}\omega^{\mu \nu}+ \frac{1}{2} m_{\omega}^2 \omega_\mu \omega^\mu - 
\frac{1}{4} \vec \rho_{\mu \nu}.\vec \rho^{\mu \nu} \nonumber \\
& + & \frac{1}{2} m_\rho^2 \vec \rho_{\mu}. \vec \rho^{\mu} -\frac{1}{3}bm_{n}(g_{\sigma}\sigma)^{3}-
\frac{1}{4}c(g_{\sigma}\sigma)^{4} +\frac{1}{4}d(\omega_{\mu}\omega^{\mu})^2 \nonumber \\
& + & \sum_{L} \bar{\psi}_{L}    [ i \gamma_{\mu}  \partial^{\mu}  - m_{L} ]\psi_{L}.
\end{eqnarray}
The leptons $L$ are assumed to be non-interacting, whereas the baryons are coupled to the meson 
(scalar $\sigma$, isoscalar-vector $\omega_\mu$ and isovector-vector $\rho_\mu$).
The constant parameters of the model are determined by fitting the nuclear matter saturation properties.
The model, however, fails to explain the experimentally observed strong $\Lambda \Lambda$ 
attraction. Mishustin \& Schaffner \cite{schaffner} corrected the model 
by the adding two mesons, the isoscalar-scalar $\sigma^*$ and 
the isovector-vector $\phi$, which couple only with the hyperons. 
In our calculation we have used two different parameter sets. The one with relatively softer EOS 
is known as TM1 parametrization and the other which generates much stiffer EOS is given by PL-Z parametrization.
The details of the parametrization can be found in Refs. \cite{schaffner,ritam2012b}.

Maintaining charge neutrality and beta equilibrium, the energy density and pressure are 
given by \cite{ritam2012b}
\begin{eqnarray}
\varepsilon & = & \frac{1}{2} m_{\omega}^2 \omega_0^2
+ \frac{1}{2} m_{\rho}^2 \rho_0^2 + \frac{1}{2} m_{\sigma}^2 \sigma^2
+ \frac{1}{2} m_{\sigma^*}^2 \sigma^{*2} + \frac{1}{2} m_{\phi}^2 \phi_0^2
+\frac{3}{4}d\omega_0^4+ U(\sigma) \nonumber \\
& & \mbox{} + \sum_b \varepsilon_b + \sum_l \varepsilon_l  \,, \\
P & = & \sum_i \mu_i n_i - \varepsilon .
\end{eqnarray}

We employ the simple MIT bag model \cite{chodos} to describe the quark EOS. The
up, down and strange quark masses are taken to be $5$MeV, $10$MeV and $200$MeV, respectively.
For the bag model, the energy density and pressure can be written as \cite{ritam2012b}
\begin{eqnarray}
\epsilon^Q &=& \sum_{i=u,d,s} 
\frac{g_i}{2 \pi^2} \int_0^{p_F^i} dp p^2\sqrt{m_i^2 + p^2}+ 
B_G\,,\label{edec}\\ 
P^Q &=& \sum_{i=u,d,s} \frac{g_i}{6\pi^2} 
\int_0^{p_F^i} dp \frac{p^4}{\sqrt{m_i^2 + p^2}}- B_G\,, 
\label{pdec}
\end{eqnarray}
where the Fermi momentum $p_F^i$ is given as $p_F^i=\sqrt{\mu_i^2-m_i^2}$. $g_i$ is the degeneracy  
of quarks $i$. $B_G$ is the bag constant and is considered a free parameter in the model.

The matter is assumed to be neutrino-free ($\mu_{\nu_e} = \mu_{\overline \nu_e} = 0$ ),
and like the hadronic matter is charge neutral and in beta equilibrium.

\section{Inclusion of magnetic field in EOS}

The magnetic field is assumed to be along the $z$ axis, and can be written as ${\vec B}=B\hat{k}$
\cite{monika,ritam2011,ritam2012}, and so the charged particles
get Landau quantized in the perpendicular direction to the field. 
Therefore, the energy of the $n$th Landau level ($n$ being the quantum number)  
\cite{chakrabarty97,yuan,broderick,chen,wei,lan} takes the form
\begin{equation}
E_i=\sqrt{{p_i}^2+{m_i}^2+|q_i|B(2n+s+1)}\,.
\end{equation}
In this equation $s$ is the spin of the particle and is equal to $s=\pm1$ for the up(+) and down(-), 
respectively. $p_i$ is the momentum along the field direction of particle $i$. 
Writing $2n+s+1=2\nu$, the energy of the particle can be rewritten as
\begin{equation}
E_i  =  \sqrt{{p_i}^2+{m_i}^{2}+2\nu |q_i|B} \nonumber \\
= \sqrt{{p_i}^2+{\widetilde{m}_{i,\nu}}^2} \,. 
\end{equation}

The number density and the energy density of the charged particles get modified, but the neutral 
particles remain unaffected. The number density and energy density of the charged particles is given by 
\begin{equation}
n_i= \frac{|q_i| B}{2 \pi^2} \sum_{\nu}
p_{f,\nu}^i \,,
\label{nmax}
\end{equation}
and 
\begin{equation}
\varepsilon_i= \frac{|q_i| B}{4 \pi^2} \sum_{\nu}
\left[ E_f^i p_{f,\nu}^i + \widetilde{m}_{\nu}^{i~2} \ln
\left( \left|
\frac{E_f^i + p_{f,\nu}^i}{\widetilde{m}^i_{\nu}} \right|
\right) \right] \,.
\end{equation}
$p_{f,\nu}^i$ is the Fermi momentum for the level with the
principal quantum number $n$ and spin $s$ and is given by
\begin{equation}
p_{f,\nu}^{i~2} =  E_f^{i~2} - \widetilde{m}_{\nu}^{i~2} \,.
\end{equation}

Therefore, the total energy density of the hadronic matter now takes the form
\begin{eqnarray}
\varepsilon & = & \frac{1}{2} m_{\omega}^2 \omega_0^2
+ \frac{1}{2} m_{\rho}^2 \rho_0^2 + \frac{1}{2} m_{\sigma}^2 \sigma^2
+ \frac{1}{2} m_{\sigma^*}^2 \sigma^{*2} + \frac{1}{2} m_{\phi}^2 \phi_0^2
+\frac{3}{4}d\omega_0^4+ U(\sigma) \nonumber \\
& & \mbox{} + \sum_b \varepsilon_b + \sum_l \varepsilon_l + \frac{{B}^2}{8 \pi^2} \,,
\end{eqnarray}
where the last term being the magnetic field contribution. The pressure can simply be represented as 
\begin{eqnarray}
P= \sum_i \mu_i n_i - \varepsilon \,.
\end{eqnarray}

For the quark matter, the thermodynamic potential in presence of strong magnetic field 
is written as \cite{chakrabarty-sahu,ritam2012}
\begin{eqnarray}
\Omega_i&=&-\frac{2g_i|q_i|B}{4\pi^2}\sum_{\nu}\int_{\sqrt{m_i^2+2\nu |q_i|
B}}^{\mu}
dE_i\sqrt{E_i^2-m_i^2-2\nu |q_i|B}.
\label {eq:om}
\end{eqnarray}

The total energy density and pressure of the strange quark matter are given by
\begin{eqnarray}
\varepsilon &=& \sum_{i}\Omega_i +B_G +\sum_{i}n_i \mu_i \nonumber \\
p&=&-\sum_i\Omega_i-B_G.
\end{eqnarray}

\section{EOS for the HS}

We use Glendenning conjecture \cite{glen} to determine the HS.
The range of baryon density, where both the hadron 
and quark phases coexists called mixed phase. The stellar matter is charge neutral,
therefore, the hadron and quark phases may be separately charged, but 
maintaining charge neutrality in the mixed phase. 
There are two approaches to construct the HS. One is the Gibbs 
construction and the other is the Maxwell construction.

The Gibbs condition determines the mechanical and chemical equilibrium 
between two phases, and is given by \cite{ritam2012,ritam2012b}
\begin{equation}
P_{\rm {HP}}(\mu_e, \mu_n) =P_{\rm{QP}}(\mu_e, \mu_n) = P_{\rm {MP}}. 
\label{gibbs}
\end{equation}
The volume fraction occupied by quark matter in the mixed phase is $\chi$, and the 
hadronic matter is $(1-\chi)$ and is connected as 
\begin{equation}
\chi \rho_c^{\rm{QP}} + (1 - \chi) \rho_c^{\rm{HP}} = 0,
\label{e:chi}
\end{equation}
where $\rho_c$ is respective charge density. This corresponds to the assumption that the 
surface tension in quark matter is almost zero. For this case the pressure varies 
continuously with energy density. The mixed phase lies in between the hadronic and quark 
phase, where both the phase coexist maintaining global charge neutrality.  

The Maxwell construction also demands 
\begin{equation}
P_{\rm {HP}}(\mu_e, \mu_n) =P_{\rm{QP}}(\mu_e, \mu_n) = P_{\rm {MP}}, 
\label{maxwell}
\end{equation}
but for this case the electron chemical potential is not constant across the boundary. 
This construction corresponds to high surface tension in quark matter. For the Maxwell 
construction there is a sudden jump in energy density from the hadronic phase to quark phase. 
There is no mixed phase in between.

\section{Results}

To begin with, we would assume that a NS, with a sudden density fluctuation at the core, undergoes a
phase transition. The phase transition is brought about by a transition front travelling from the centre 
to the surface of the star. Previous calculation of the energy release estimation assumed that the baryonic mass of the 
star remains constant. The number of baryons before and after the conversion does not change.
This assumption may not be valid for every scenario as there may be some matter ejection from the star. 
For the phase transition induced collapse of NS to QS, 
there may be a shock development at the centre of the star which gives enough momentum for the 
matter to eject from the outer layers of the star \cite{woosley,harko}.
It was suggested by Fryer \& Woosley \cite{woosley} that the mass ejection may be as high as $0.2$ times 
solar mass. Recent calculation also shows that the phase transition (only deconfinement) in the NS can 
proceed via a deflagration or a detonation \cite{bhat1,herzog} with such high velocities that 
the whole star is converted to two flavor quark matter within a few millisecond. For such cases 
the velocity is so high that there ought to be some matter ejection from the outer layers. If such is 
the case, the baryon mass conservation condition cannot be maintained strictly. 
We assume that the matter is ejected before the phase transition front 
reaches the outer layers. Therefore, the ejected matter is normal hadronic matter and not quark matter. 

First we would calculate the energy
release with no matter ejection and then we would calculate the energy release when there is a matter ejection of
$M_B=0.2$ solar mass. The resultant baryonic mass of the NS after
matter ejection is denoted as $M_{BE}$. The conversion may continue up to the surface or may die down
after some distance. This depends on the initial energy difference between the matter phases at the 
centre of the star and also on the initial density and spin fluctuation of the star \cite{glen}. If the conversion 
process continues to the surface we have a SS, and if it stops in between we have a HS. For the HS we have assumed both
Gibbs (Gib) and Maxwell (Max) prescription.

The relativistic mean field EOS model for nuclear matter is used to construct the NS. 
The final star may be a SS or a HS. We construct the 
HS based on Glendenning construction. In our calculation we have used two different parametrization for the 
hadronic EOS, and had regulated the quark EOS by changing the bag constant. The masses of the light quarks 
are quite bounded and we take them to be $5$MeV (up) and $10$MeV (down). The mass of strange (s) quark is 
still not well established, but expected to lie between $100-300$MeV. Therefore, we have kept the mass of 
the s-quark fixed at $200$MeV. We take different bag constant ($B_G$) to have different quark matter EOS.
For simplicity, we will denote ${B_G}^{1/4}=160$ MeV$=B_g$. 

\begin{figure}
\centering
\begin{tabular}{cc}
\begin{minipage}{200pt}
\includegraphics[width=200pt]{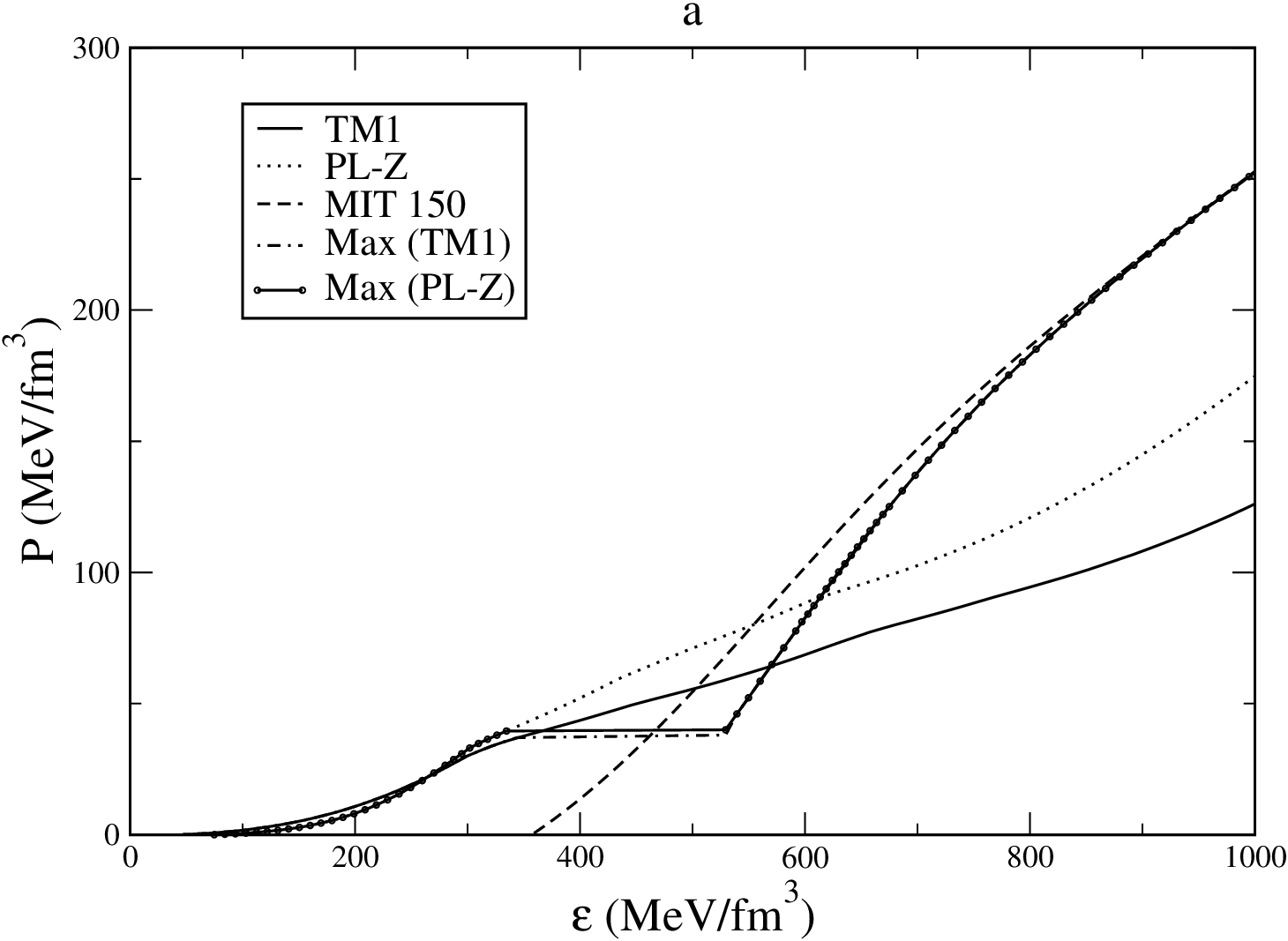} 
\end{minipage}
\begin{minipage}{200pt}
\includegraphics[width=200pt]{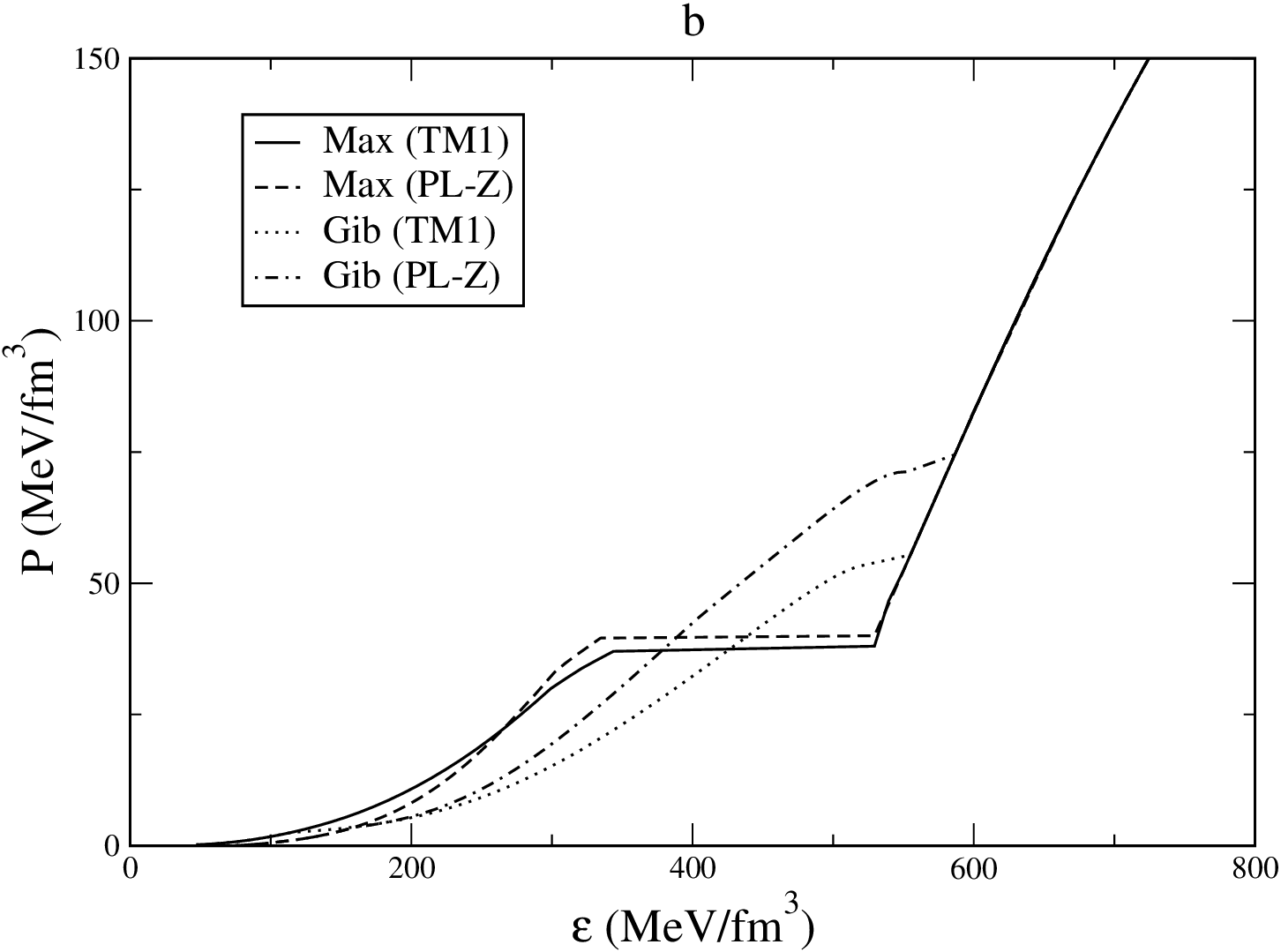} 
\end{minipage} \\
\end{tabular}
\caption{In (a) pressure as a function of energy density is plotted for different EOS. The hadronic EOS are denoted as TM1 
and PL-Z, the strange EOS with bag pressure $150$ MeV as MIT 150, 
and the mixed EOS as Max (hadronic and quark parametrization are specified). In (b) The difference between
Max and Gib construction are explicitly shown.}
\label{fig1}
\end{figure}

There are several studies to model the density dependence of $B_g$ and we  
choose one of the widely used model given in refs. \cite{adami,blaschke}. 
Without going into much detail, which can be found in refs. \cite{ritam2012,ritam2012b}, we give the expression
for the density variation of the bag constant 
\begin{eqnarray}
B_{gn}(n_b)  =  B_\infty  +  (B_g  -  B_\infty)  \exp  \left[  -\beta  \Big(
\frac{n_b}{n_0} \Big)^2 \right] \:, 
\label{bag}
\end{eqnarray}
where $B_\infty=130$ MeV is the lowest value of the bag constant which it attains at asymptotically 
high densities (known from experiments).
The bag pressure mentioned in the paper is the initial value of bag pressure ($B_g$ in the equation). 
As the density increases the bag pressure decreases and reaches $130$MeV asymptotically, 
the decrease rate is controlled by $\beta$. We take $\beta=0.003$ (which can generate massive QS \cite{ritam2012b}).

However, a recent calculation by MingFeng et al. \cite{ming} shows that for a varying bag constant in the 
quark EOS, an extra term gets added in the matter pressure. The matter pressure is now given by
\begin{equation}
p=-\sum_i\Omega_i-B_G+n_b\frac{\partial B_G}{\partial n_b},
\label{mterm}
\end{equation}
where, $n_b$ is the number density. The last term makes the quark EOS 
softer. As the EOS is softer the maximum mass also decreases. In this work we have followed the 
prescription for calculating the EOS of quark matter.

\begin{figure}
\centering
\begin{tabular}{cc}
\begin{minipage}{200pt}
\includegraphics[width=200pt]{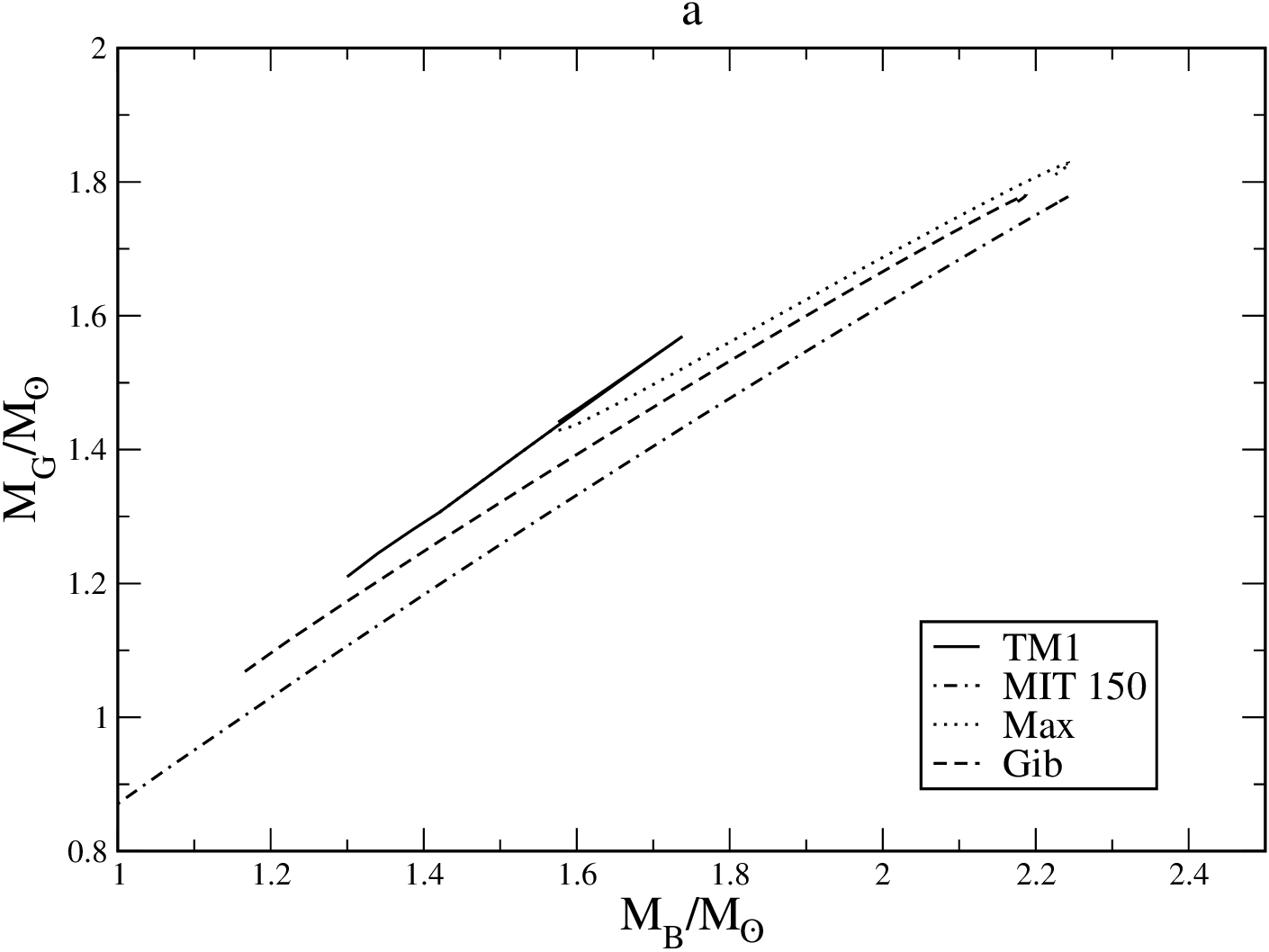} 
\end{minipage}
\begin{minipage}{200pt}
\includegraphics[width=200pt]{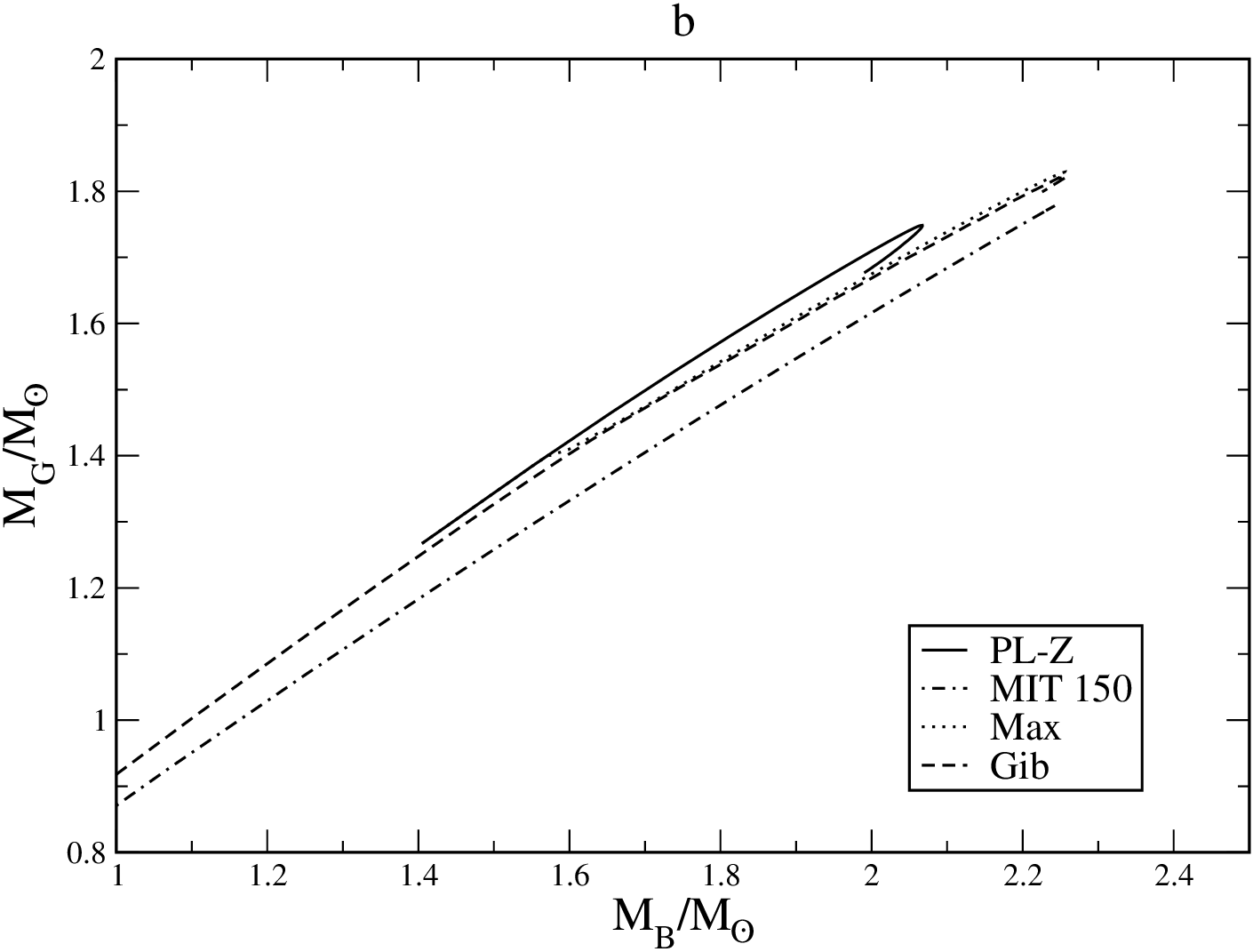} 
\end{minipage} \\
\end{tabular}
\caption{Gravitational masses as a function of baryonic masses are plotted for different EOS. (a) is for TM1 parametrization 
(hadronic and hybrid) and (b) for Pl-Z parametrization.}
\label{fig2}
\end{figure}

The strange matter is absolutely stable if the bag pressure is $150$ MeV. Therefore, for such bag constant
the resultant star is a strange star (SS). However, if the bag pressure $160$ MeV, the strange matter is not absolutely
stable and we get a hybrid star. The hybrid star can have a mixed phase (denoted as Gib) or there can be a direct jump
from quark to hadronic matter (denoted as Max). In this work, the bag pressure for the strange star is always $150$ MeV and for the 
HS the bag pressure is $160$ MeV.

In Fig. 1a, we have plotted pressure as a function of energy density for different EOS.
The hadronic EOS with TM1 and PL-Z parametrization. The figure shows that the EOS with TM1 parametrization 
is softer than the PL-Z parametrized EOS. The mixed EOS curves has been plotted with $B_g=160$ MeV, 
with the above mentioned hadronic parametrizations. The difference between the Gibbs and Maxwell prescription for the 
mixed EOS is shown in Fig. 1b. The EOS with Gibbs prescription shows an extended mixed phase region 
whereas the EOS with Maxwell prescription shows a jump in the energy density 
(corresponding to a jump in baryon density) from the hadronic to quark matter.
For the Maxwell construction with TM1 parametrization there is a jump in the 
nuclear density from $0.34 fm^{-3}$ to $0.56 fm^{-3}$ and for PL-Z the jump is $0.36 fm^{-3}$ to $0.57 fm^{-3}$. 
For the Gibbs construction with TM1 parametrization 
the mixed phase occurs in the range of $0.13 fm^{-3}-0.56 fm^{-3}$ and for PL-Z the range is $0.16 fm^{-3}-0.57 fm^{-3}$.

\begin{table}
\begin{center}
\begin{tabular}{|l|l|l|l|l|l|l|l|l|}
\hline
$NS \rightarrow SS$ & $M_B$ & $M_G(NS)$ & $M_P(NS)$ & $M_G(SS)$ & $M_P(SS)$ & $E_G$ & $E_I$ & $E_T$ \\ \hline
$TM1L \rightarrow MIT150$ & $1.71$ & $1.546$ & $1.781$ & $1.41$ & $1.677$ & $0.57$ & $1.86$ & $2.43$ \\
$PL-Z \rightarrow MIT150$ & $1.945$ & $1.67$ & $1.957$ & $1.573$ & $1.915$ & $0.66$ & $1.07$ & $1.73$ \\ \hline
$NS \rightarrow HS$ \\ \hline
$TM1L \rightarrow Max$ & $1.71$ & $1.546$ & $1.781$ & $1.501$ & $1.752$ & $0.28$ & $0.52$ & $0.80$ \\ 
$TM1L \rightarrow Gib$ & $1.71$ & $1.546$ & $1.781$ & $1.468$ & $1.747$ & $0.78$ & $0.61$ & $1.39$ \\ 
$PL-Z \rightarrow Max$ & $1.945$ & $1.67$ & $1.957$ & $1.638$ & $1.975$ & $0.89$ & $-0.32$ & $0.57$ \\ 
$PL-Z \rightarrow Gib$ & $1.945$ & $1.67$ & $1.957$ & $1.676$ & $1.983$ & $0.79$ & $-0.08$ & $0.71$ \\ \hline
\end{tabular}
\caption{Table showing the energy released during the phase transition from NS to SS/HS for the 
non magnetic star. There is no matter ejection during the conversion. The masses are 
in terms of solar masses and the energies in terms of $10^{53}$ergs.}
\label{table1}
\end{center}
\end{table}

We assume that there is no matter ejection from the outer layers of the star during the PT. 
In Table \ref{table1}, we have
given the corresponding proper mass and gravitational mass of the NS and QS (SS/HS) for the same baryonic mass ($M_B$).
For the hadronic star we have used two different
model parametrization which gives different $M_B$ for the NS. For the TM1 model $M_B$ is $1.71 M_{\odot}$ and for PL-Z
model it is $1.945 M_{\odot}$. The phase transition occurs and the NS is converted to QS. The resultant 
$M_B$ for the QS is the same that of the initial NS. If the phase transition happens throughout 
the whole star and the star is converted to a SS.
The energy liberated is $2.43 \times 10^{53}$ ergs. If the phase transition stops in between and we have a 
resultant HS the liberated energy is somewhat less. The HS without a mixed phase region liberates the minimum energy
and is $7-8 \times10^{52}$ ergs. 

To explain the energy release, we plot gravitational mass against baryonic mass curves for different EOS (fig 2). The energy liberated is due to the
difference in the gravitational mass between a NS and a QS for a particular baryonic mass. This depends on the relative stiffness of the
curves (this stiffness is not same as the EOS curve stiffness). We find that the most stiff curve is for the hadronic EOS and 
the least stiff is the strange EOS. The Max curve is slightly stiffer than the Gib curve.
Although the strange EOS is the stiffer than hadronic EOS, the NS with hadronic EOS reaches its 
maximum mass at much lower central energy density. So in the baryonic mass vs. gravitational mass plot the NS stiffness is the highest and the 
SS stiffness is the lowest. Similar argument follows for the Max and Gib curves (as Gib curve has extended quark region in the mixed phase). 
Therefore, the energy liberated during the conversion of NS to Max HS is the least and is the maximum for 
the conversion of NS to SS. What physically happens is that, for the conversion of NS to SS the conversion takes place throughout 
the star and so the energy liberated is larger, whereas for the HS the conversion takes place only up to a certain region.
The energy released during the conversion of Max HS is less than that of Gib HS, This is because for the Gib HS the conversion process 
also continues in the mixed phase whereas the Max HS has no mixed phase so the conversion process stops much earlier.

\begin{table}
\begin{center}
\begin{tabular}{|l|l|l|l|l|l|l|l|l|l|}
\hline
$NS \rightarrow SS$ & $M_B$ & $M_{BE}$ & $M_G(NS)$ & $M_P(NS)$ & $M_G(SS)$ & $M_P(SS)$ & $E_G$ & $E_I$ & $E_T$ \\ \hline
$TM1L \rightarrow MIT150$ & $1.71$ & $1.51$ & $1.379$ & $1.555$ & $1.265$ & $1.474$ & $0.59$ & $1.45$ & $2.04$ \\
$PL-Z \rightarrow MIT150$ & $1.945$ & $1.745$ & $1.53$ & $1.76$ & $1.445$ & $1.712$ & $0.75$ & $0.86$ & $1.61$ \\ \hline
$NS \rightarrow HS$ \\ \hline
$TM1L \rightarrow Max$ & $1.71$ & $1.51$ & $1.379$ & $1.555$ & $1.379$ & $1.555$ & $0$ & $0$ & $0$ \\ 
$TM1L \rightarrow Gib$ & $1.71$ & $1.51$ & $1.379$ & $1.555$ & $1.325$ & $1.544$ & $0.76$ & $0.2$ & $0.96$ \\ 
$PL-Z \rightarrow Max$ & $1.945$ & $1.745$ & $1.53$ & $1.76$ & $1.502$ & $1.768$ & $0.64$ & $-0.14$ & $0.50$ \\ 
$PL-Z \rightarrow Gib$ & $1.945$ & $1.745$ & $1.53$ & $1.76$ & $1.449$ & $1.757$ & $0.5$ & $0.5$ & $0.55$ \\ \hline
\end{tabular}
\caption{Table showing the energy released during the phase transition from NS to SS/HS for the 
non magnetic star. The initial baryonic mass of the star is the same as that of Table \ref{table1}. 
There is matter ejection from the outer layer due to shock bounce, and is of baryonic 
mass $0.2$ solar mass.}
\label{table2}
\end{center}
\end{table}

Table \ref{table2} shows the energy liberated during the conversion of NS to SS/HS, when there is matter ejection from the 
surface layers. The amount of matter ejected is taken to be $0.2 M_{\odot}$ of $M_B$. Therefore, the resulting baryonic mass 
denoted as $M_{BE}$ is $(M_B -0.2) M_{\odot}$. As the $M_B$ decreases the gravitational mass and 
proper mass of the NS and QS also decreases. We find that the energy released during such conversion is slightly less than
that from the previous case. This is because as the star loses matter it becomes less massive and so the energy liberated is 
slightly less. 

\begin{figure}
\centering
\includegraphics[width=3.3in]{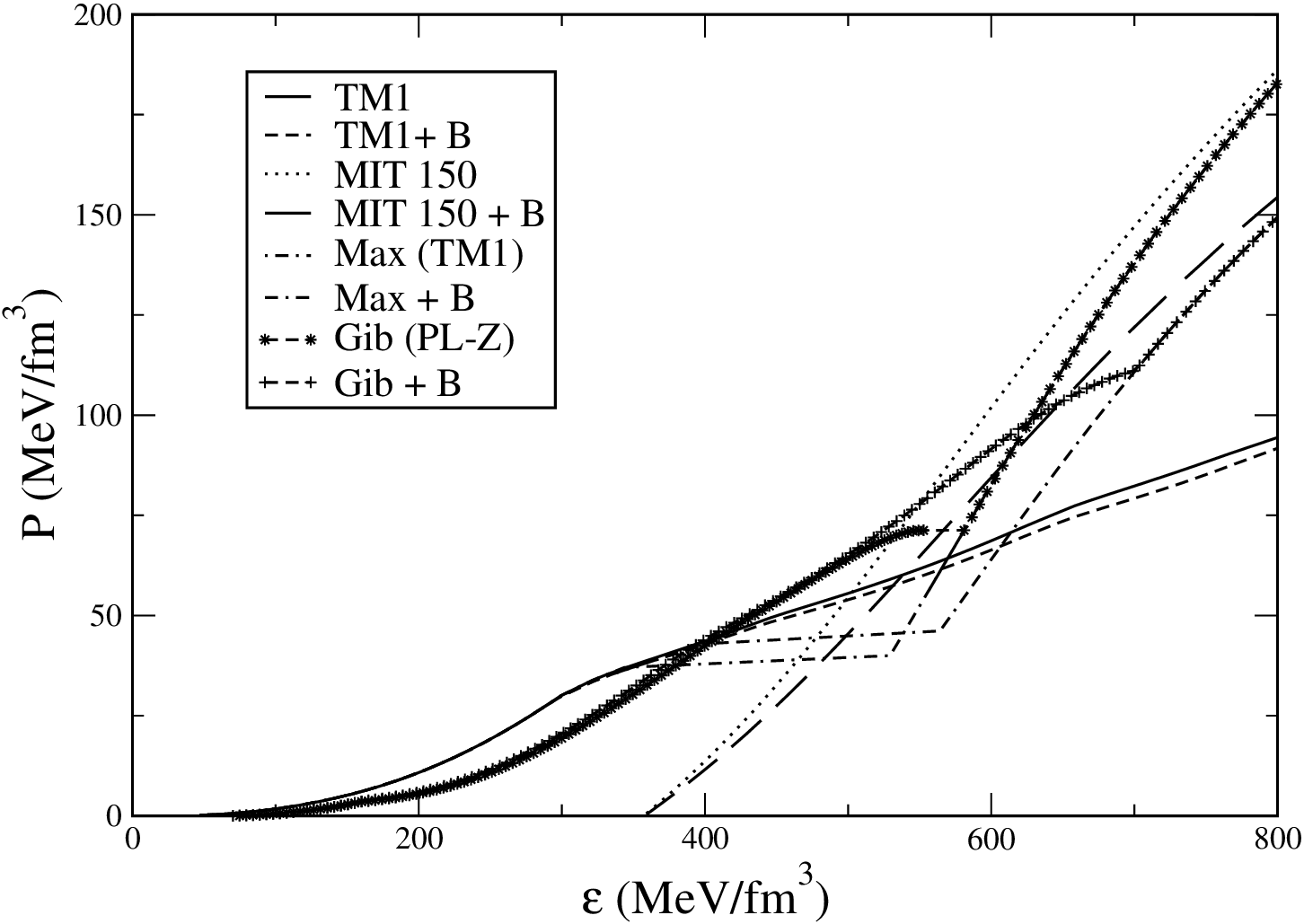}
\caption{Pressure vs. energy density plot for different EOS with and without magnetic field.
The magnetic field is parametrized according to that given in Eqn. \ref{mag-vary}.
The hadronic EOS is given as TM1, and its magnetic counterpart is denoted as TM1+B. The same representation 
is adopted for the strange and mixed EOS. The straight curve is the hadronic (TM1) EOS and the short dash is its magnetic counterpart.
The dotted curve is for the MIT 150 and long dash is for its magnetic counterpart. The broken curves are for the Max TM1 
(dash-dot the original and dash-dash-dot the magnetic) and the curves with star (original) and plus (magnetic) are for the Gib PL-Z.}
\label{fig3}
\end{figure}

Next we considered the effect of magnetic field in the EOS.
The surface field strength observed in magnetars lie between $10^{14}$G to $10^{15}$G. 
It is usually assumed that the magnetic field increases as one goes towards the centre of the 
star, and a simple calculation involving flux conservation of the progenitor star 
yields the central magnetic field as high as $\sim 10^{17}-10^{18}$G. Without going 
into much detail we follow the prescription of the field variation given in Refs. \cite{chakrabarty97,monika}
\begin{equation}
{B}(n_b)={B}_s+B_0\left\{1-e^{-\alpha \left(
\frac {n_b}{n_0} \right)^\gamma}\right\},
\label{mag-vary}
\end{equation}
where $\alpha$ and $\gamma$ determines the magnetic field variation across the star for fixed
surface field $B_s$ and central field $B_0$. The value of $B$ depends primarily on $B_0$.
We keep the surface field strength fixed at $B_s=10^{14}$G. We put $\gamma=2$ and $\alpha=0.01$ 
to have the field variation. It was shown that magnetic with field strength greater than few times $10^{18}$G,
makes the star unstable. Therefore, for our calculation we assume a conservative value of the central 
field to be $10^{17}$G. The variation in the $\alpha$ and $\gamma$ would give interesting result as 
far as the magnetic field variation in the star is concerned, but for our calculation of the energy release
the results would not change much. So we assume a fixed $\alpha$ and $\gamma$ values, $\alpha=0.01$ and $\gamma=2$.

In Fig. \ref{fig3} we have plotted magnetic field induced EOS curves. 
Qualitatively the magnetic induced EOS curves are much softer than the 
non magnetic curve. This is because the magnetic pressure due to Landau quantization act in the 
opposite direction of the matter pressure. Also, the magnetic stress acts
towards the matter energy density. These two collectively reduces the stiffness of the magnetic field induced EOS.
The curves shows that the magnetic field has very little effect on hadronic region (low-density regime) as the 
field strength there is low. As we go towards the centre of the star the magnetic field strength increases and therefore
the effect on the mixed and quark phase are higher. The quark sector (high-density regime) is the most affected and the 
mixed phase is moderately affected (intermediate regime). 
The mixed phase region also gets extended due to the magnetic field 
and the region now is $0.13 fm^{-3}-0.65 fm^{-3}$ for the TM1 parametrization and $0.17 fm^{-3}-0.69 fm^{-3}$ for PL-Z parametrization.
The jump in the nuclear density from the nuclear to quark matter for the Maxwell construction is from 
$0.39 fm^{-3}$ to $0.57 fm^{-3}$ for the TM1 parametrization and from $0.43 fm^{-3}$ to $0.59 fm^{-3}$ for the PL-Z parametrization.

\begin{figure}
\centering
\begin{tabular}{cc}
\begin{minipage}{200pt}
\includegraphics[width=200pt]{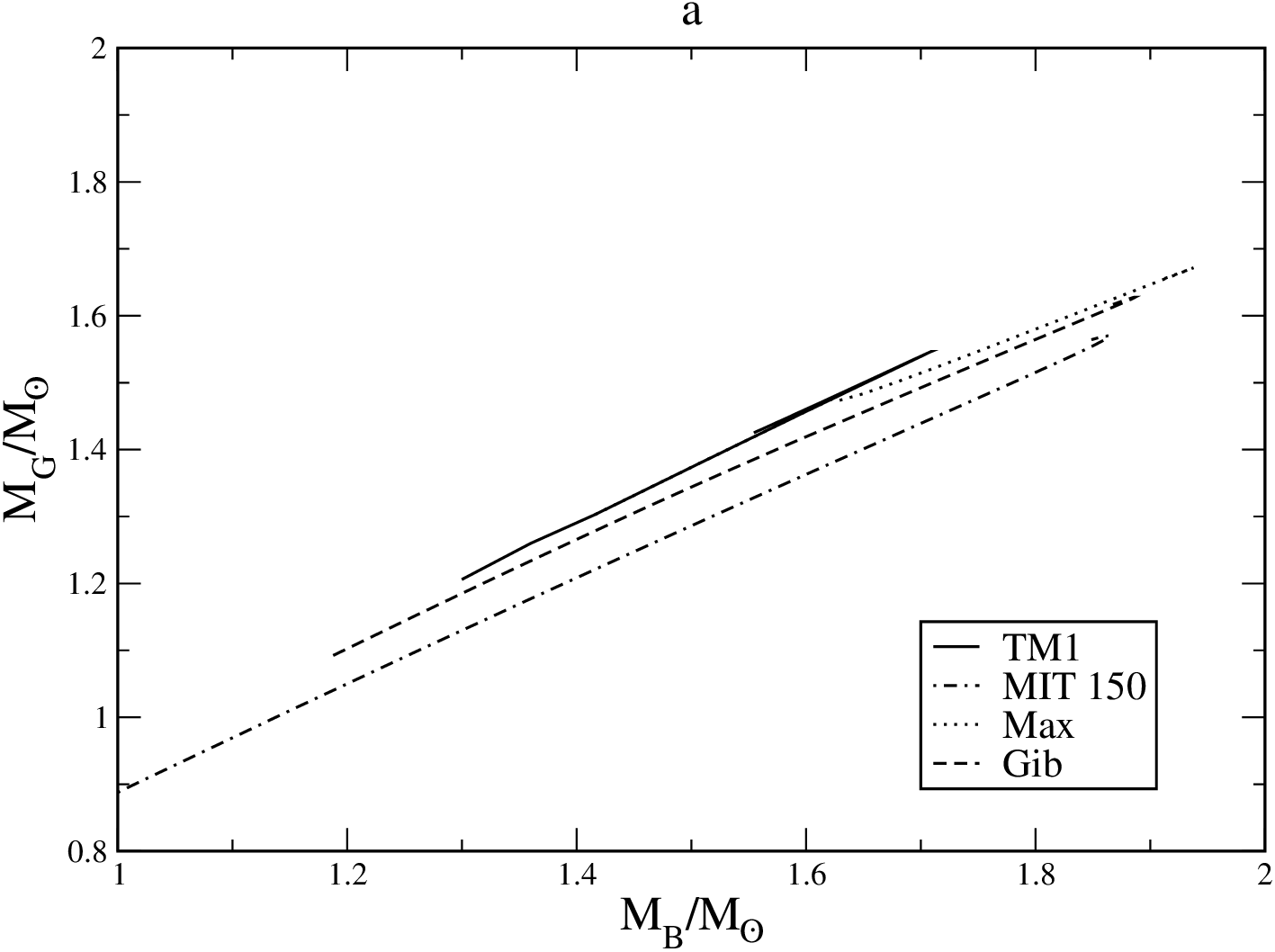} 
\end{minipage}
\begin{minipage}{200pt}
\includegraphics[width=200pt]{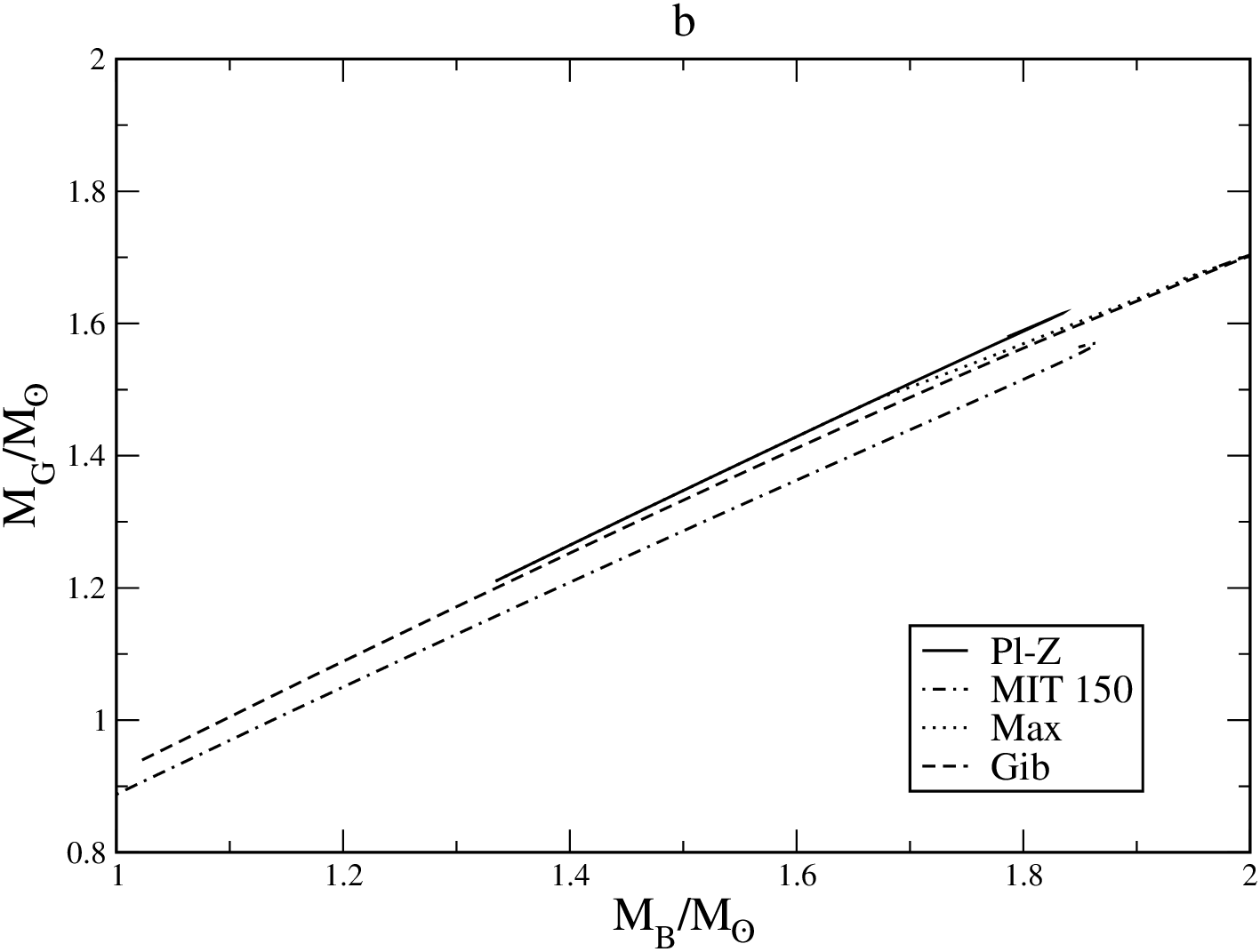} 
\end{minipage} \\
\end{tabular}
\caption{Gravitational masses as a function of baryonic masses are plotted with for magnetars. (a) is with TM1 parametrization
and (b) is with PL-Z parametrization.}
\label{fig4}
\end{figure}

Table \ref{table3} gives the energy releases during a phase transition of NM to a quark magnetar 
(QM), which can be either be SM or a HM. The magnetic field strength and 
configuration is mentioned above. We first 
assume that there is no matter ejection. As the magnetic field makes the EOS softer, both the baryonic mass and gravitational mass
decreases. As the star becomes less massive (see fig \ref{fig4}), the energy liberated is much less.

Table \ref{table4} shows the energy liberated during the conversion of NM to SM/HM, when there is matter ejection 
from the surface layers. We again assume the amount of matter ejected to be $0.2 M_{\odot}$ of $M_B$. 
For this case the star is less massive, therefore the 
energy released is also less. It is interesting to note that for some conversion from NS to Max HS the 
energy liberated is zero as the NS GM and the SS GM becomes same for a particular BM. Physically it means that the PT happens
without any energy release.

\begin{table}
\begin{center}
\begin{tabular}{|l|l|l|l|l|l|l|l|l|}
\hline
$NS \rightarrow SS$ & $M_B$ & $M_G(NS)$ & $M_P(NS)$ & $M_G(SS)$ & $M_P(SS)$ & $E_G$ & $E_I$ & $E_T$ \\ \hline
$TM1L \rightarrow MIT150$ & $1.7$ & $1.536$ & $1.772$ & $1.436$ & $1.728$ & $1.0$ & $0.79$ & $1.79$ \\
$PL-Z \rightarrow MIT150$ & $1.825$ & $1.606$ & $1.889$ & $1.532$ & $1.877$ & $1.11$ & $0.21$ & $1.32$ \\ \hline
$NS \rightarrow HS$ \\ \hline
$TM1L \rightarrow Max$ & $1.7$ & $1.536$ & $1.772$ & $1.512$ & $1.767$ & $0.34$ & $0.09$ & $0.43$ \\ 
$TM1L \rightarrow Gib$ & $1.7$ & $1.546$ & $1.772$ & $1.49$ & $1.777$ & $0.9$ & $-0.09$ & $0.81$ \\ 
$PL-Z \rightarrow Max$ & $1.825$ & $1.606$ & $1.889$ & $1.583$ & $1.885$ & $0.34$ & $0.07$ & $0.41$ \\ 
$PL-Z \rightarrow Gib$ & $1.825$ & $1.606$ & $1.89$ & $1.576$ & $1.875$ & $0.29$ & $0.25$ & $0.54$ \\ \hline
\end{tabular}
\caption{Table showing the energy released during the phase transition from NM to QM (SM/HM). 
The magnetic field is parametrized according to that given in Eqn. \ref{mag-vary}. 
There is no matter ejection during the conversion.} 
\label{table3}
\end{center}
\end{table}

The energy liberated during the conversion of NS to SS is of the amount of $10^{53}$ ergs, and it can account for 
the cosmological origin of the GRB. Such huge energy release is only comparable to the energy observed during
the GRB. The detailed picture of the conversion mechanism and the central engine for the GRB is still not 
well understood. The above described process can be responsible for the energetic of 
the GRB. In this paper, we have concentrated more on the difference in gravitational binding energy due to
phase transition which is accompanied by matter ejection from the outer layers. It should be realized that for 
a realistic star with rotation and magnetic field, there would be asymmetric matter ejection and also 
asymmetric energy deposition. The amount of matter ejected, the
fraction of energy which spend on heating the star and the fraction which is spend on
propagating the conversion front is still not clear. However, if a significant amount of energy 
released goes into the energetic of electron-positron pairs production, it can account for the GRB energy.

The energy released during the conversion of NM to HM is of the order of $10^{52}$ ergs, and are 
therefore difficult to power GRB at cosmological distances. However, the magnetars would release energy more efficiently 
from their magnetic poles than the normal NS. Therefore, in magnetars the energy liberated from the star would have a large
Lorentz number and can account for the giant flare activity.

\begin{table}
\begin{center}
\begin{tabular}{|l|l|l|l|l|l|l|l|l|l|}
\hline
$NS \rightarrow SS$ & $M_B$ & $M_{BE}$ & $M_G(NS)$ & $M_P(NS)$ & $M_G(SS)$ & $M_P(SS)$ & $E_G$ & $E_I$ & $E_T$ \\ \hline
$TM1L \rightarrow MIT150$ & $1.7$ & $1.5$ & $1.371$ & $1.545$ & $1.285$ & $1.509$ & $0.9$ & $0.64$ & $1.54$ \\
$PL-Z \rightarrow MIT150$ & $1.825$ & $1.625$ & $1.445$ & $1.655$ & $1.38$ & $1.645$ & $0.98$ & $0.18$ & $1.16$ \\ \hline
$NS \rightarrow HS$ \\ \hline
$TM1L \rightarrow Max$ & $1.7$ & $1.5$ & $1.371$ & $1.545$ & $1.371$ & $1.545$ & $0$ & $0$ & $0$ \\ 
$TM1L \rightarrow Gib$ & $1.7$ & $1.5$ & $1.371$ & $1.545$ & $1.342$ & $1.556$ & $0.72$ & $-0.2$ & $0.52$ \\ 
$PL-Z \rightarrow Max$ & $1.825$ & $1.625$ & $1.445$ & $1.655$ & $1.445$ & $1.655$ & $0$ & $0$ & $0$ \\ 
$PL-Z \rightarrow Gib$ & $1.825$ & $1.625$ & $1.445$ & $1.655$ & $1.428$ & $1.657$ & $0.26$ & $0.04$ & $0.30$ \\ \hline
\end{tabular}
\caption{Table showing the energy released during the phase transition from NM to QM. 
The magnetic field is parametrized according to that given in Eqn. \ref{mag-vary}. 
The initial baryonic mass of the star is the same as that of Table \ref{table3}. 
There is matter ejection from the outer layers of baryonic mass $0.2$ solar mass.}
\label{table4}
\end{center}
\end{table}

This mechanism of energy production and connecting them with GRB is not unanimously accepted.
Some models predicts that it is sufficient to power GRB \cite{ma1}, but
other model predicts that although the energy may be high, but to produce high Lorentz factor the energy
goes down to $10^{46}-10^{49}$ergs \cite{woosley}. However, in a recent
paper by Cheng et al. \cite{harko} argued that due to the difference of gravitational binding 
energy of the star after the phase transition some matter near the stellar surface would be ejected.
This matter would further be accelerated by electron-positron pairs created by neutrino-antineutrino
annihilation. The mass ejection could give rise to high Lorentz factor and also high luminosity
needed for GRB.

We should mention that the EOS which was used for the calculation are softer due to the presence of 
strangeness (hyperons in hadronic EOS and strange quarks in quark EOS) in them. The maximum mass that the EOS can produce
is much less than $2$ solar mass, which is the new limit of NS mass from precise measurement of pulsars 
J1614-2230 and J0348+0432. However, for the hadronic EOS, if we consider non hyperonic EOS with 
TM1 parametrization the maximum mass $2.18$ solar mass. For the PL-Z parametrization the same is $2.3$ solar mass, 
both of which can easily satisfy the mass constraint. New calculation points that
with strong vector-meson coupling in SU(3) symmetry \cite{weiss1,weiss2}, even hyperonic star can generate massive NS.

In the quark sector, even with density dependent bag constant in the MIT bag model the maximum mass of QS are below $2$ solar 
mass. This is due to the presence of extra term coming from consideration of density dependent bag constant (Eq. 26).
The EOS of quark sector is still much debated and new model (NJL and PNJL) are being used to satisfy the strong 
interacting behaviour. However, still there is no exact calculation and the EOS are very parameter dependent.
In our model if we consider the effect from quark interaction and colour superconducting matter as used by Alford et al.
and Weissenborn et al. \cite{alford,weiss3}, the maximum mass of SS becomes $2.05$ solar mass and that of HS
becomes $2.13$ solar mass (with $a4=0.67$ for quark interaction and $\Delta=100$ MeV for gap energy), both of which 
satisfy the new mass limit. However, with such EOS the qualitative nature of our results remains the same. Even 
with such consideration, the hadronic EOS is still the steepest and the quark EOS is the softest. Therefore, our conclusion that 
the energy liberated during the conversion of NS to SS is greater than that liberated during the conversion of NS to
HS holds (there is only quantitative change in the values by $20-30\%$). The magnetic field makes all the EOS softer. 
Therefore the observation that the energy liberated during conversion of magnetars is less than that liberated during phase 
transition in normal pulsars also remains unaltered.

\section{Summary and Conclusion}

In this work, we have calculated the energy released during the conversion of NS to QS (SS/HS).
The total energy released is the sum of the gravitational and internal energy of conversion. 
Our first assumption was that the conversion starts with a sudden spin down of the star, resulting in a
huge density fluctuation at the core, initiating the phase transition. 
We also assume that some matter is ejected from the outer layers of the star due to several shock bounce, 
and therefore, the change in baryonic mass.  
The conversion may continue up to the surface or may 
die out after some distance. This depends on the energy difference between the matter phases at the 
centre of the star and also on the initial density and spin fluctuation of the star.
The final star may be a SS or a HS. 
For the NS we have considered relativistic mean field EOS model of hadronic matter. 
For the quark matter (SS/HS), we have considered simple MIT bag model. We construct the 
HS based on Glendenning construction. In our calculation we have used two different parametrization for the 
hadronic EOS, and had regulated the quark EOS by changing the bag constant.

First we have shown the energy liberated during the conversion of NS to QS, with no matter ejection.
With a fixed baryonic mass, the energy liberated is obtained from the difference in the gravitational
mass of the initial NS and final QS. For such a case the liberated energy is always close to $10^{53}$ ergs.

Next, we had shown the energy liberated during the conversion of NS to QS, with matter ejection from the surface layer
of the NS. The amount of matter ejected is taken to be $0.2 M_{\odot}$ of $M_B$. Therefore, the resulting baryonic mass 
denoted as $M_{BE}$ is $(M_B -0.2) M_{\odot}$. Due to matter ejection as the $M_B$ decreases the gravitational mass and 
proper mass of the NS and QS also decreases. The energy liberated during the conversion of NS to QS with matter ejection 
is always less than that of the energy liberated during the conversion of NS to QS with no matter ejection.
 We have also studied the conversion of NS to SS/HS having high surface magnetic field (observed magnetars).
We have denoted them as NM and QM (SM/HM). The energy liberated during the conversion of magnetars is less than that for 
normal pulsars. 

One thing that clearly points out is that the conversion of NS to HS liberates energy few times less than 
that for the conversion of NS to SS. Therefore, it would be difficult for them to power GRB from cosmological distances.
For the Max HS the energy liberated is the lowest and sometimes it is zero, which indicates the PT there occurs
without any observable signal.

The energy liberated during the conversion of NS to QS is of $10^{53}$ ergs, and it can account for 
the cosmological origin of the GRB if the Lorentz factor is high. 
The energy released during the conversion of NM to QM is of the order of $10^{52}$ ergs, and are 
therefore difficult to power GRB at cosmological distances. However, the magnetars would release energy more efficiently 
from their magnetic poles than the normal NS. Therefore, in magnetars the energy liberated from the star with large
Lorentz number and can account for the giant flares activity. 

The detailed picture of the conversion mechanism and the central engine for the GRB is still not 
well understood. However, the above described processes are thought to be responsible for the energetic of 
the GRB. In this paper, we have concentrated more on the difference in gravitational binding energy due to
phase transition which is accompanied by matter ejection from the outer layers. We have seen the difference in 
the energy release for normal pulsars and magnetars. The amount of matter ejected, the
fraction of energy which goes towards heating of the star and the fraction spent on 
maintaining the conversion would ultimately determine the fate of energy release. 
However, if a significant amount of energy released goes into the energetic of 
electron-positron pairs production, it can account for the GRB energy. 

Assuming the baryonic mass of the star to remain unaltered, Bombaci and Dutta \cite{bombaci} performed 
their calculation and found that the NS has always greater mass than the final
SS. Neutrino-antineutrino pair annihilation to electron-positron 
pairs deposit energy of the order of $10^{49}$ergs \cite{cheng}. Neutron proton scattering by neutrinos 
inside the dense star deposit energy at least two to three order further higher. Due to the shock of the 
phase transition some matter near the stellar surface may be ejected and would further be accelerated by 
electron-positron pairs and may eventually oscillate. This may give rise to high Lorentz factor and high luminosity
needed for GRB. These two are the most efficient processes to account for the energetic of the GRB.
As the actual process takes place at cosmological distances, we do not 
have a better understanding of such phenomena. As there are no other source of external 
energy available to the star, other
than the rotational energy, a huge amount of energy may be used in the actual conversion.
We can only conclude that if a small fraction of the energy of the conversion is released it may 
manifest itself at least in the form of giant flares, which are usually associated with the magnetars. 
However, if a substantial
amount of energy is released, it may account for the energies observed during the GRB. Therefore, more detailed studies in 
theoretical and observational front is still needed for the pulsars and magnetars to have a better understanding
of the energetic of the conversion.

On the other hand, with new observation of $2$ solar mass pulsar gives very strong constraints for the EOS 
describing compact stars. The EOS used in our calculation cannot satisfy the new mass limit. However, this
can be remedied by considering new prescriptions for both hadronic and quark matter EOS. With the improved
EOS, the basic results of our calculation does not have any qualitative change but only small quantitative 
change. With new techniques of precise mass measurement, the physics of pulsars is 
entering a new phase. As the exact nature of strong interactions is still far from being settled, the 
quark EOS being very model dependent, the maximum mass of NS and QS calculated from theoretical consideration are going to 
evolve further in future. To get a clear picture of energy release and its detailed mechanism we need more
detailed microscopic studies and this work is the first step towards that direction.

{}
\end{document}